\documentstyle[12pt]{article}

\renewcommand{\thefootnote}{\fnsymbol{footnote}}

\begin{document}
\thispagestyle{empty}

\begin{center}
{\large{\bf Algebra of potentials of the  volume-preserving
      vector fields.}} \vspace{1.5cm}\\

{\large  R.L. Mkrtchyan \footnote{E-MAIL: mrl@dircom.erphy.
armenia.su}}\vspace{0.5cm}\\

{\it Yerevan Physics Institute}\\
{\it Alikhanyan Br. st.2, Yerevan, 375036 Armenia }
\end{center}

\begin{abstract}
The algebra of volume-preserving vector fields is considered. The
potentials for that fields are introduced, and induced algebra
of potentials is considered. It is shown, that this algebra fails to
satisfy the Jacoby identity. Analogy with hamiltonian mechanics is
developed,
as well as 3-cocycle interpretation of corresponding expressions.
\end{abstract}

\vfill
\setcounter{page}0
\renewcommand{\thefootnote}{\arabic{footnote}}
\setcounter{footnote}0

\newpage

\newcommand{\ra}{\rightarrow}

\pagestyle{plain}

\section{\bf Introduction.}

  The Lie algebras of vector fields on some n-dimensional manifold
M, preserving the volume form $\omega$ on that manifold, have found some
applications in physics recently \cite {witt,avan,broo,seli,niss,arat,manv}.
They appear for $n=2$ (i.e. for the case
of area-preserving transformations) as an algebra of hidden symmetries
of $d=1$ closed string field theory \cite{witt,avan}, and the same
$d=1$ string theory,
at a certain radius of compactification, possess
an $n=3$ volume-preserving
diffeomorphisms
algebra. The algebras
of vector fields, of which the algebra of volume-preserving fields is the
subalgebra, prove their importance in physics and mathematics \cite{fukh}
in many important cases, and deserve further study.

         The main aim of this letter is to construct an algebra of
potentials for a divergenseless vector fields, i.e. the fields, preserving
some volume form, and to show, that this algebra
unavoidably fails to satisfy the Jacoby identity, so it is not a Lie algebra.
The construction is carried out first in Sect.2 for three-dimensional case,
when that potentials turn out to be one-forms on that manifolds, and
is generalized to an arbitrary $n$ in Sect.5.
  This construction has a very close analogy with the construction
of the algebra of hamiltonian vector fields, when one considers the subalgebra
of vector fields, which maintain the symplectic structure, i.e. some
nondegenerate closed two-form. The corresponding potentials are the usual
Hamiltonians, which now are the scalar functions on the manifold.That
analogy is considered in Sect.3.

  The other point of view on the objects, appeared in Sec.2,
is that we obtained a 3-cocycle of the algebra of volume-preserving
vector fields, with values in the exact one forms. That approach will be
discussed in Sect.4.

  Some possible developments of these ideas are described in the
Conclusion:  the   hamiltonian   mechanics   generalization,   the
generalization
of the notion of the algebra of the symmetries of the theories, etc.

\newpage

\section{\bf Construction of the algebra}

        Let's consider a three-dimensional manifold, which we shall
take as $R^3$, (this is not a restriction, since our considerations will be
local), with the volume
3-form, which can be brought to the simple form $\epsilon_{\mu\nu\lambda}
dx^\mu \wedge dx^\nu \wedge dx^\lambda $.
The vector fields
$\xi^\mu$,
which maintain that form, are characterized by the property
\begin{equation}
\partial_\mu\xi^\mu=0
\label{1}
\end{equation}
which means, that they can be parameterized by the
vector-potentials $A_\mu$ as
\begin{equation}
\xi^\mu \equiv \xi^\mu (A)=\epsilon^{\mu\nu\lambda} \partial_\nu A_\lambda
\label{2}
\end{equation}
This  parametrization  is  degenerate  in  a  sense,   that   many
potentials correspond
to the same vector fields: the gauge transformed potential
\begin{equation}
A_\mu + \partial _\mu \varphi
\label{3}
\end{equation}
leads to the same vector field $\xi^\mu$.

Evidently, divergenseless vector fields form the Lie algebra, since
  the commutator of such a fields gives again divergenseless field - from
\begin{equation}
\bigl\lbrack\xi _1 , \xi _2 \bigr\rbrack= \xi _3
\label{4}
\end{equation}
follows that $\xi^\mu$ satisfies the eq. (\ref{1})
\begin{equation}
\partial_\mu\xi^\mu _3=0
\label{5}
\end{equation}

Let's denote through $A, B$ and $C$ the potentials for the vector fields
$\xi^\mu _i=0 , i=1,2,3$, so $\xi^\mu_1= \xi^\mu (A), \xi^\mu_2= \xi^\mu (B),
\xi^\mu_3= \xi^\mu (C)$.
$C$ is defined by $A$ and $B$ only up to the  gauge transformation. The
calculation shows, that $C$ is given by

\begin{equation}
 C_\mu=\epsilon_{\mu\nu\lambda}\xi^\nu _1 \xi^\lambda _2
 + \partial_\mu\phi
\label{6}
\end{equation}
where $\phi$ is an arbitrary function, which cannot be obtained from the
commutation relation between vector fields. So, one can try to
introduce an algebra of potentials, with the following binary operation
("Poisson bracket", see next section)
\begin{equation}
\bigl\lbrack A , B \bigr\rbrack _\mu = C _\mu
\label{7}
\end{equation}
where $C$ has to be given by (\ref{6}) with some concrete choice for $\phi$.
$\phi$ may depend on
$A$ and $B$, and may satisfy some additional requirements. The natural
requirements, aimed to keep the algebra of potentials as close as
possible to the original algebra of vector fields, are the requirement
of linearity over $A, B$, and the  antisymmetry under the interchange of $A$
and
$B$.
The possible choice, which also has the property of being invariant
with respect to the independent gauge transformations of $A$ and $B$, is
$\phi=0.$
Let's accept this definition now, the more general choice will be
discussed later.

        The real problem arises when one considers the Jacoby identity
for this bracket. Since the algebra of potentials simulate  the
algebra of vector fields up to a gauge transformations, nothing
guarantee
now that the rhs of Jacoby identity will be exactly zero, and not the pure
gauge. The straightforward calculation shows

\begin{equation}
\bigl\lbrack A ,\lbrack B,C \bigr\rbrack\rbrack_\sigma + (cycl. perm.)=
\partial _\sigma \epsilon _{\mu\nu\lambda}\xi^\mu (A)\xi^\nu(B)\xi^\lambda(C)
\label{8}
\end{equation}
where $A,B,C$ are three independent vector-potentials. We observe the
appearance
of the violation of the Jacoby identity, by the pure gauge terms.

  The non-trivial question is whether one can remove the rhs of the Jacoby
identity by the appropriate choice of $\phi$ in the commutation relation
(\ref{6}).
It is easy to understand, that the only possible choice for  $\phi$, which
gives in the rhs of Jacoby identity terms, similar to (\ref{8}), is
\begin{equation}
\phi= k\epsilon^{\mu\nu\lambda}\partial _\mu (
A _\nu B _\lambda )
\label{9}
\end{equation}
where $k$ is an arbitrary constant.
This choice maintains the antisymmetry and linearity, but violates gauge
invariance.
With this new bracket the Jacoby identity looks like
\begin{eqnarray}
& &  \bigl \lbrack A ,\lbrack B,C \bigr\rbrack\rbrack_\sigma + (cycl. perm.)=
\partial_\sigma\epsilon _{\mu\nu\lambda}\xi^\mu(A)\xi^\nu(B)\xi^\lambda(C)
+ \nonumber\\
& &(k^2+k)\partial _\sigma \partial _\mu \lbrack
\xi^\mu (A) \xi^\lambda (B) C _\lambda + (cycl. perm.)\rbrack
\label{10}
\end{eqnarray}
The last term is gauge noninvariant (changes on $ (k^2+k)\partial
_\sigma \partial _\mu \lbrack
\xi^\mu (A) \xi^\lambda (B) \partial _\lambda \sigma + (cycl. perm.)\rbrack
)$ under gauge transformation of (e.g.)$ C: C \rightarrow C + \partial
\sigma$,
 hence, cannot cancel first, gauge-invariant, term.

So, we conclude, that it is impossible to restore the Jacoby identity
using the freedom in definition (\ref{6}), by the appropriate choice of $\phi$
The most natural choice  is $\phi=0$, since it has an additional property
of gauge invariance.
        Thus, we end up with an algebra of the vector-potentials with the
binary operation
\begin{equation}
\bigl\lbrack A , B \bigr\rbrack_\mu=\epsilon_{\mu\nu\lambda}\xi^\nu
(A) \xi^\lambda (B)
\label{11}
\end{equation}
which is antisymmetric, linear and gauge-invariant w.r.to the gauge
transformations of A and B, but does not satisfy the Jacoby identity.

\section{\bf Hamiltonian analogy}

Let's consider the usual hamiltonian formalism. One starts
from some closed two-form $\omega= \omega _{\mu\nu} dx ^\mu \wedge dx ^\nu$
instead of closed three-form of previous
section, and consider the vector fields, which maintain that form - the
hamiltonian vector fields. This is the requirement
\begin{equation}
\partial _\mu \omega_{\lambda\nu}\xi^\nu - (\mu \leftrightarrow \nu) =0
\label{12}
\end{equation}
instead of (\ref{6}), and so $\xi^\mu$ can be parameterized as

\begin{equation}
\xi^\mu \equiv \xi^\mu (H) = \omega^{\mu\nu}\partial _\nu H
\label{13}
\end{equation}
where $H$ is the analog of $A$, but now is a scalar function. The analog of
gauge
transformations (\ref{3}) is the shift of $H$ on some arbitrary constant
\begin{equation}
H \rightarrow H + const
\label{14}
\end{equation}
so, $H$ can be recovered from (\ref{13}) only up to a constant.
The commutator of two hamiltonian vector fields gives another hamiltonian
vector field, and denoting the corresponding Hamiltonians as $H_1, H_2, H_3$,
we obtain
\begin{equation}
H_3 =\lbrace H_1 ,H_2 \rbrace + c(H_1 ,H_2)
\label{15}
\end{equation}
where the bracket in the r.h.s. is the usual Poisson bracket, and
$c$ is a constant, which depends on $H_1, H_2$ and may satisfy some
constraints.
This expression is an analog of (\ref{7}). Imposing the natural conditions of
antisymmetry and linearity, one can try also to fulfill the Jacoby
identity, maintaining in that way the Lie algebra nature of hamiltonian
vector fields. One easily finds, that this last condition leads
to the equation on $c(H_1 ,H_2)$ which is an equation on two-cocycle of the
algebra of
hamiltonian vector fields (more extensive discussion of cocycles
equations see below):
\begin{equation}
c(H_1 ,\lbrace H_2,H_3\rbrace) + (cycl. perm.) =0
\label{16}
\end{equation}
The trivial solution $c=0$
leads to a usual Poisson-Lie algebra structure on the space of Hamiltonians,
solutions with $c\not= 0$ give a central extensions of that algebra.

\section {\bf  The cocycle interpretation.}

  The expressions of preceding Section have an interpretation on
the language of the cocycles of the Lie algebra of the volume-preserving
vector fields. The notion of cocycles of the (gauge) Lie groups, was used
intensively in the study of anomalies \cite{fadd}, where 2-cocycles appeared,
and also possible appearance and physical meaning of 3-cocycles
were discussed \cite{jackiw, grossm}. In this section we shall interpret the
rhs
of Jacoby
identity of Sect. 2 as a 3-cocycle of the group of a volume-preserving
diffeomorphisms.

The cochains of the Lie algebra X are an antisymmetric
polylinear functionals on that algebra with values in some X-modules.
The coboundary operation on the space of cochains is given by the equation
\begin{eqnarray}
\delta c_n (x_1,...,x_{n+1})&=&\sum _{k<l} (-1)^{k+l}
c_n ([x_k,x_l],x_1,...,\hat {x_k},...,\hat {x_l},...,x_n)+\nonumber\\
& & \sum_{k=1}^{n+1} x_k (-1)^k c_n (x_1,...,\hat {x_k} ,...,x_{n+1})
\label{17}
\end{eqnarray}
where hat on the argument means that it is absent.

For application to our case, we choose as basic Lie algebra the algebra
of volume-preserving vector fields, and a module is a space of exact one-forms
with trivial action of vector fields.
Due to the construction (\ref{8}), rhs of that equation gives
the closed 3-cochain $c_3(\xi_1,\xi_2,\xi_3)=\partial_\sigma
\epsilon_{\mu\nu\lambda}(\xi_1^\mu,\xi_2^\nu,\xi_3^\lambda)$:

\begin{equation}
\delta c_3 (x_1,...,x_4)=0
\label{18}
\end{equation}
The nontriviality of that cocycle (i.e. whether it can be represented as
the coboundary of some 2-cochain) requires more rigorous approach and will
be discussed elsewhere.

\section {\bf Higher-dimensional generalization.}

        Higher-dimensional generalization of the algebra of potentials of
volume-preserving transformations of three-dimensional space, introduced
in Sect.2, is straightforward. The same constraint (\ref{1}) on
volume-preserving vector fields now has a solution parameterized in
$d$-dimensional space by $(d-2)$-th rank tensors $A_{\mu_1 ... \mu_{d-2}}$:
\begin{equation}
\xi^\mu \equiv \xi^\mu (A)=\epsilon^{\mu\mu_1 ... \mu_{d-1}} \partial_
{\mu_1}
A_{\mu_2 ... \mu_{d-1}}
\label{19}
\end{equation}

Induced algebra structure on the space of potentials is given by the
evident generalization of (\ref{7}):
\begin{equation}
\bigl\lbrack A , B \bigr\rbrack _{\mu_1 ... \mu_{d-2}} =
C _{\mu_1 ... \mu_{d-2}}
\label{20}
\end{equation}

where (d-2)-form C is given by

\begin{equation}
C _{\mu_1 ... \mu_{d-2}}=\epsilon_{\mu_1 ... \mu_d}\xi^{\mu_{d-1}} \xi^{\mu_d}
 + \partial_{\lbrack\mu_1}\phi_{\mu_2 ... \mu_{d-2}\rbrack}
\label{21}
\end{equation}
where brackets mean antisymmetrization over indexes inside them.
We see, that now function $\phi$ of (\ref{6}) becomes a $(d-3)$-form, according
to the fact, that gauge transformation of potentials, which doesn't affect
the corresponding vector fields, adds to the potentials an
exterior derivative of $(d-3)$-form.

The main statement of Sect.2 of impossibility of fulfillness the Jacoby
identity remains unchanged also in this case. Namely, the Jacoby identity
now looks like
\begin{equation}
\bigl\lbrack A ,\lbrack B,C \bigr\rbrack\rbrack_{\mu_1 ... \mu_{d-2}} +
(cycl. perm.)=
\partial_{\lbrack\mu_1}\epsilon _{\mu_2 ... \mu_{d-2}\rbrack\nu\lambda\sigma}
\xi^\nu(A) \xi^\lambda (B) \xi^ \sigma (C)
\label{22}
\end{equation}
Again, one cannot remove the rhs of Jacoby identity by changing the
definition of bracket, using an arbitrariness in definition of C in
(\ref{21}).

\section{\bf Conclusion.}
In previous sections we introduce an unconstrained potentials for a
volume-preserving vector fields, and, in complete analogy with construction
of Poisson bracket, construct a similar bracket on the space of potentials.
The main difference is in that new bracket doesn't satisfy the Jacoby
identity. Construction is carried out in all dimensions.

The other interpretation of corresponding expressions is as a 3-cocycles
of the Lie algebra of volume-preserving vector fields.
This construction can be viewed as a tool
for obtaining the cocycles of the algebra of volume-preserving vector
fields, and one may try to generalize that to other subgroups of algebra
of all vector fields.

The interesting direction of development follows from the analogy,
discussed in Sect.3. Namely, it is possible to consider the "equations
of motion" of one-forms, in the form, analogous to usual hamiltonian
mechanics

\begin{equation}
{d\over{dt}}  A = \lbrace H, A \rbrace
\label{23}
\end{equation}
where H is a given "Hamiltonian" one-form, and "the motion" means,
that an arbitrary one-forms $A$ change in time according to the equation
(\ref{23}).
In usual hamiltonian mechanics the same equation may be interpreted as
a change of function according to the change of its arguments, so the
whole mechanics consists from motion of point in the phase space. In
our case such an interpretation is not straightforward, but nevertheless, more
general point of view is possible, and one can consider different motions
in the space of one-forms, induced by different Hamiltonian one-forms H.
It is interesting to find and classify finite-dimensional orbits of
such a motions.
Actually, according to \cite{priv} in the three-dimensional case the
present construction, in the special case of the 1-form H, given by
the expression $ H=FdG $, gives the Namby's odd-dimensional mechanics
\cite{namb}.
Finally, we would like to mention another question, arising in connection
with the algebras with
Jacoby identity, violated by pure gauge terms.
Since the observables in gauge theories are gauge invariant, one
can try to study theories with non-Lie-algebraic type of symmetries,
the deviation from the Lie algebras being  purely  gauge,  and
unessential
on the level of observables. Another direction of investigation
is the consideration of coadjoint action of the algebra of
potentials.

I'm grateful to R.Hayrapetian, D.Karakhanyan, H.Khudaverdian and R.Manvelyan
for a discussions of the
present work.

\end{document}